\documentclass[aps,prl,showpacs,twocolumn,amsmath,amssymb]{revtex4}
\usepackage{graphicx}

\newcommand{\Hh}{ {\hat H} }
\newcommand{\ww}{ {\hat w} }
\newcommand{\uu}{ {\hat u} }
\newcommand{\ii}{ {\rm i} }
\newcommand{\ee}{ {\rm e} }

\newcommand{\dd}{ \mathrm{d} }
\newcommand{\aaa}{ {\hat\alpha} }

\newcommand{\xbra}[1]{{}_\sharp{\langle #1 \vert}}

\newcommand{\xket}[1]{{\vert #1 \rangle}_\sharp}

\begin{document}
\title{Operator Space Entanglement Entropy in XY Spin Chains}
\date{\today}
\author{Iztok Pi\v{z}orn}
\author{Toma\v{z} Prosen}
\affiliation{Department of Physics, FMF, University of Ljubljana, 
            Jadranska 19, SI-1000 Ljubljana, Slovenia}


\begin{abstract}
The complexity of representation of operators in quantum mechanics can 
be characterized by the operator space entanglement entropy (OSEE). 
We show that in the homogeneous Heisenberg XY spin $1/2$ chains
the OSEE for initial local operators grows at most logarithmically with time. 
The prefactor in front of the logarithm generally depends only 
on the number of stationary points of the quasi-particle dispersion relation
and for the XY model changes from $1/3$ to $2/3$ exactly at the point of
quantum phase transition to long-range magnetic correlations in the
non-equilibrium steady state. In addition, we show that the presence
of a small disorder triggers a saturation of the OSEE.
\end{abstract}

\maketitle


Complexity of many-body quantum systems has two remarkable manifestations: on one hand, certain strongly correlated (entangled) many-body states (e.g. cluster states \cite{cluster}) can be used to perform universal quantum computation
and thus yield exponential gain over best known classical algorithms for 
certain tasks.
On the other hand, weak entanglement (finite range of quantum correlation) is necessary for
quantum states in order to be efficiently described classically \cite{vidal}. 
The latter fact is the reason
behind the success of numerical methods, such as {\em density matrix renormalization group} (DMRG) \cite{white,vidal} which approximate quantum states by their most entangled components, e.g. as matrix product states (MPS) where the number of components depends on the quantum entanglement of the state. While the
ground states of one-dimensional many-body systems are weakly 
entangled \cite{hastings,verstraeteprb73,latorre}
and can be efficiently computed by the DMRG, the time evolution of 
generic quantum states produces entanglement which results in a growing
need of resources (see e.g. \cite{prosenpre75}). Such algorithms are only efficient if the computational 
costs grow at most polynomially with time or, equivalently, 
the entanglement entropy 
which is a measure of quantum entanglement, 
grows no faster than logarithmically 
-- which is the case only for very particular initial states and model systems. 

Another option is to simulate time-dependent operators in Heisenberg picture 
using the same principles where the operators are represented as products of 
matrices, i.e. matrix product operators (MPO) \cite{verstraeteprl93}.
This approach can be efficient even when the simulation of (pure) quantum states is 
{\em not} 
as is the case in transverse Ising spin chain
\cite{calabreseJSM4,prosenpre75,pleniohartmann,znidaricpra}.
In this context the relevant quantity determining the efficiency
is the \emph{operator space entanglement entropy} (OSEE) \cite{prosenpra76}
which is an operator-space analogue of the usual bipartite entanglement entropy 
and presents a complementary quantification to other measures of entanglement in 
operator space \cite{zanardi}.
As in the evolution of quantum states, similar limitations also apply
to operators which accounts to the inefficiency of time-evolution of 
generic operators attributed by the exponential growth of resources in time.
Nevertheless, 
it was shown \cite{prosenpre75,prosenpra76} that \emph{local} operators 
-- on the contrary -- 
can always be simulated efficiently
in the \emph{integrable} transverse Ising model with the OSEE being either finite 
(increasing logarithmically) for initial operators which are represented repectively 
as products of a finite (infinite), number of Majorana fermions 
[having finite (infinite) index].

In this letter we shall consider OSEE of local Heisenberg operators of
infinite index in the quantum XY spin-$1/2$ chain, or any translationally invariant 
spin chain which can be solved by Wigner-Jordan transformation.
We show that OSEE in such models generally increases logarithmically in time, where the prefactor
is given universally as $1/6$ times the {\em number of stationary points of the quasi-particle
dispersion relation}. We identify two generic regimes in the XY model with prefactors $1/3$ and $2/3$ 
exactly corresponing to two quantum phases of a non-equilibrium steady state 
of open XY spin chain far from equilibrium \cite{prosenprl101} and the regime of the gapless XY model
with a prefactor $1/6$.
Investigating the effect of disorder we find, as an interesting consequence of Anderson-like localization in operator space, that OSEE saturates in time even for the infinite index initial operators.


Dynamics of the quantum XY spin $1/2$ chain of length $n$ is described in terms of 
Pauli operators $\sigma_j^{\mathrm{x},\mathrm{y},\mathrm{z}}$, $j\in\{1,2,\ldots,n\}$ by the Hamiltonian
\begin{equation}
H = \sum_{j=1}^{n-1}\bigg(\frac{1+\gamma}{2} \sigma_j^{\rm x}\sigma_{j+1}^{\rm x} + 
\frac{1-\gamma}{2} \sigma_j^{\rm y}\sigma_{j+1}^{\rm y} \bigg)
  + \sum_{j=1}^{n} h \sigma_j^{\rm z},
\label{eq:HXY}
\end{equation}
which is conveniently expressed as a quadratic form 
$H = \underline{w}\cdot \mathbf{H} \underline{w}$ in terms of $2n$
Hermitian \emph{Majorana operators} 
\begin{equation}
w_{2j-1} = \big(\prod_{l<j}\sigma_l^{\rm z} \big) \sigma_j^{\rm x},
\quad
w_{2j} = \big(\prod_{l<j}\sigma_l^{\rm z} \big) \sigma_j^{\rm y},
\end{equation}
obeying the anticommutation relation $\{w_j,w_l\}=2\delta_{jl}$, 
and $2n\times 2n$ antisymmetric Hermitian matrix $\mathbf{H}$.
For the XY-model (\ref{eq:HXY}) the only upper-diagonal elements of $\mathbf{H}$ read 
$H_{2j,2j+1} = -(\ii/2) \frac{1+\gamma}{2}$, 
$H_{2j-1,2j+2} = (\ii/2) \frac{1-\gamma}{2}$, 
$H_{2j-1,2j} = -(\ii/2) h$
for $j\in\{1,\ldots,n\}$. 

As shown in \cite{prosenpra76,prosen3Q} we identify 
$4^n$ dimensional Pauli algebra with a 
{\em Fock space of operators} describing $2n$ \emph{adjoint fermions} (a-fermions),
with an orthonormal canonical basis 
$\xket{P_{\underline\alpha}} = %
\xket{ w_1^{\alpha_1}w_2^{\alpha_2} \cdots w_{2n}^{\alpha_{2n}}}$, 
$\alpha_j \in \{0,1\}$. We define a set of adjoint 
\emph{annihilation linear maps} $\ww_j$ defined as 
$\ww_j \xket{P_{\underline\alpha}} = \alpha_j \xket{ w_j P_{\underline\alpha}}$
which satisfy the canonical anticommutation relations
$\{\ww_j^{},\ww_l^{}\}=0$,
$\{ \ww_{j}^{}, \ww_{l}^\dagger\}=\delta_{jl}$ for $j,l\in\{1,\ldots,2n\}$.
The Heisenberg dynamics in the adjoint (operator) space is 
given by a formal Schr\" odinger equation
$\ii (\dd/\dd t) A(t) = \hat{H} A(t)$,
or for the Majorana generators 
$\ii (\dd/\dd t){\underline w}(t) = 4 \mathbf{H} {\underline w}$,
which defines the adjoint hamiltonian
$\Hh = -{\rm ad\,}H \equiv [\bullet,H] = -4 \underline{\ww}^\dagger \cdot \mathbf{H} \underline{\ww}$%
.

The operator space entanglement entropy (OSEE) of an arbitrary operator
$A = \sum_{\underline{\alpha}} a_{\underline{\alpha}} P_{\underline{\alpha}}$
is defined as the bipartite entanglement entropy 
of the adjoint state (a-state)
$\xket{A} = \sum_{\underline{\alpha}}%
           a_{\underline{\alpha}}\xket{P_{\underline{\alpha}}}$
in the operator space.
The transformation between the physical basis 
$\{\sigma_1^{s_1} \cdots \sigma_{n}^{s_n}\}$ 
and the basis $\{P_{\underline{\alpha}}\}$ is
a simple permutation (with multiplications by $\pm \ii$ or $\pm 1$)
and despite the nonlocality it maps the first $n/2$ spin operators to the first
$n$ a-fermions and vice versa.
Therefore, the OSEE of initially product operator $A$ 
can be calculated by essentially following Ref. \cite{latorre} 
using the correlation matrix
$\Gamma_{jl} = \xbra{A} \ww_j^\dagger \ww_l^{} \xket{A}$ for
$j,l \in \{1,\ldots,n\}$. The correlation matrix is Hermitian 
and the OSEE is calculated from its eigenvalues $\gamma_j$ as
\begin{equation}
S =  
  \sum_{j=1}^{n} H_2(\gamma_j)
\label{eq:OSEE} 
\end{equation}
for $H_2(x)=-x \ln x - (1-x) \ln (1-x)$.

\begin{figure}[!h]
\centering
\includegraphics[width=75mm]{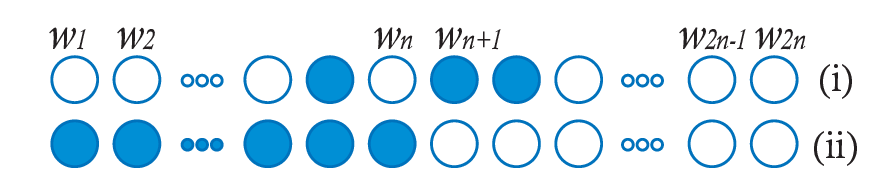}	
\caption{Schematic examples of finite (i) and infinite (ii) index operators in operator Fock space.}
\label{fig:majorana}
\end{figure}
In this work we will 
only be interested in the time evolution of initially \emph{local} operators
i.e. products of a finite number of Pauli operators $\sigma_j^{s_j}$.
This implies that, either such operator $A$ has (i) a 
\emph{finite index} in Majorana representation (see Fig. \ref{fig:majorana}) 
-- i.e. a finite number of occupied $a$-fermi states -- 
or (ii) $A$ has an \emph{infinite index} (in the limit $n\to\infty$).
It was shown analytically \cite{prosenpra76} 
for the transverse Ising chain ($\gamma=1$) 
that in the case (i) the 
OSEE saturates in time while in the case (ii) the numerical results 
give firm evidence that the OSEE grows logarithmically as 
\begin{equation}
S = c \ln t + c' 
\label{eq:logOSEE}
\end{equation}
where the coefficient $c$ is the same for any
infinite-index operator 
$A = F B$ where $F=w_1 \cdots w_n$ and
$B$ is a finite-index operator. 
Thus we will eventually consider only the simplest 
infinite-index operator $F$ corresponding to a half-filled Fermi sea of
a-fermions (Fig.~\ref{fig:majorana} case ii). 
Despite the nonlocality of the operator $F = \ii^{n/2} \sigma_1^{\rm z}\cdots \sigma_{n/2}^{\rm z}$, 
its entanglement properties are similar to those of
a local operator $\sigma_{n/2}^{\rm x} \equiv \ii^{-n/2+1} F w_{n}$.
%
%

The correlation matrix $\mathbf{\Gamma}$ for a time
dependent initially product a-state $\xket{A(t)}$ 
is calculated employing the Heisenberg picture in the operator space
using $\xbra{A(t)}\ww_j^\dagger\ww_l^{}\xket{A(t)}=%
\xbra{A}\ww_j^{\dagger}(t)\ww_l^{}(t)\xket{A}$ where 
$\ww_j(t)$ are
obtained from $(\dd/\dd t) \ww_j = -\ii [\ww_j, \Hh]$
as 
%
%
\begin{equation}
\underline{\ww}(t) = \mathbf{\Phi}^T \underline{\ww} 
\enskip \text{ where }\enskip  
\mathbf{\Phi} = \ee^{-4 \ii t \mathbf{H}}
.
\label{eq:Phi}
\end{equation}
Note that 
the matrix $\mathbf{\Phi}$ is real for any $w$-quadratic Hamiltonian
with Hermitian anti-symmetric matrix $\mathbf{H}$.
%
The correlation matrix used to calculate the OSEE (\ref{eq:OSEE}) thus reads 
$
\Gamma_{jl}(t) = 
\sum_{p=1}^{2n} \Phi_{pj}\Phi_{pl} \xbra{A}\ww_p^\dagger \ww_p \xket{A}.
$

\begin{figure}
\centering
\includegraphics[width=0.9\columnwidth]{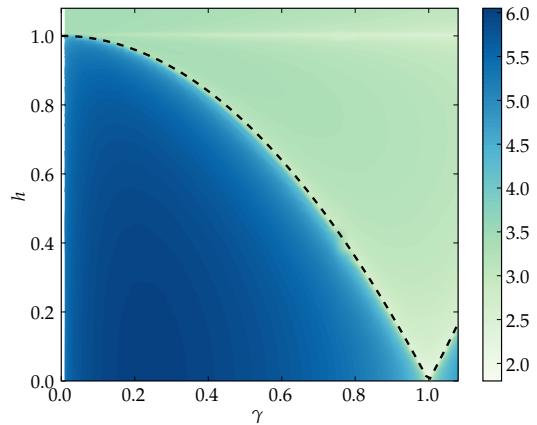}
\caption{%
   OSEE of $\xket{F(t)}$ in the XY($\gamma$,$h$) model with 
   open boundaries at $t=800$ and $n=1800$. The dashed line denotes 
   the critical line $h_{\rm c} = \vert 1-\gamma^2 \vert$.
}
\label{fig:grid}
\end{figure}
In Fig.~\ref{fig:grid}, showing phase diagram $\gamma-h$ of OSEE in the XY-model 
at fixed large time $t=800$,  two distinct regimes are identified (plus a critical regime at $h=1$).
We will later show that these two regions are bordered by 
the critical field strength $h_{\rm c} = \vert 1-\gamma^2\vert$ also observed in the {\em open}
nonequilibrium quantum XY chain \cite{prosenprl101} regardless of the way the chain
was coupled to the reservoirs. Therefore, the origin of such phenomen is attributed 
to the adjoint Hamiltonian $\Hh$ which
gives exactly the unitary part in the master equations governing the density matrix evolution  of the 
{\em boundary opened} spin chain \cite{prosen3Q}.


In the thermodynamic limit $n\to\infty$ we expect that 
the dynamics will not change significantly if
periodic boundary conditions in \emph{Majorana space} are imposed.
For finite-index initial operators being a product of Majorana operators 
near the center of the chain (Fig.~\ref{fig:majorana} case (i)) 
the OSEE indeed agrees with the open boundary result which is also
consistent with the area law as the area of the boundary remains the same.

\begin{figure}
\centering
\includegraphics[width=0.9\columnwidth]{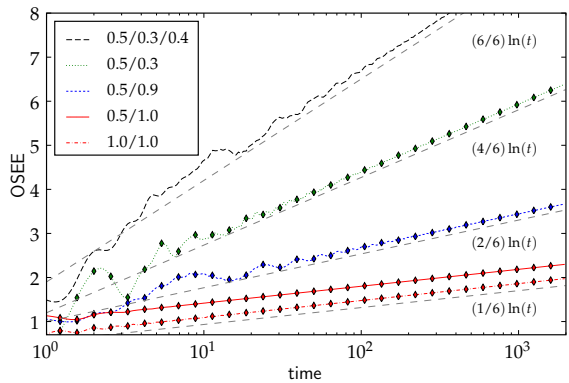} 
\caption{OSEE of $\xket{F(t)}$ for the XY($\gamma$,$h$) model (\ref{eq:HXY}) 
with open boundaries (lines) and periodic boundary conditions in Majorana 
space (OSEE divided by two, symbols).
Parameters are designated by $\gamma$/$h$.
The upper-most line denotes OSEE for the model with an 
additional term (\ref{eq:additional}) with $\mu=0.4$.
}
\label{fig:fOSEE}
\end{figure}

For infinite-index operators however the area of the boundary 
doubles when periodic boundary conditions in Majorana space are imposed
which results in a OSEE multiplied by a factor of two.
Dividing the OSEE for periodic case accordingly, we obtain 
a perfect agreement with the results for the open boundary case 
as seen in Fig.~\ref{fig:fOSEE}. The upper-most line will be described later.

The advantage of periodicity in \emph{Majorana space} 
is that $\Hh$
can be diagonalized using the Fourier expansion 
\begin{equation}
\begin{pmatrix} \ww_{2j-1} \\ \ww_{2j} \end{pmatrix} = %
\frac{1}{\sqrt{n}} \sum_{k=1}^{n} \ee^{\ii j\phi_k} %
\begin{pmatrix} \uu_{2k-1} \\ \uu_{2k} \end{pmatrix}, 
\quad \phi_k = \frac{2\pi k}{n}
\end{equation}
followed by a Bogoliubov transformation. 
First we rewrite $\Hh$ in terms of $\uu_k$ which satisfy the same
anticommutation relations as $\ww_j$, 
\begin{equation}
\Hh = \sum_{k=1}^{n}\big( a_k^{} \uu_{2k}^\dagger \uu_{2k-1}^{} + %
                          a_k^* \uu_{2k-1}^\dagger \uu_{2k}^{}\big)
\label{eq:Hh2}
\end{equation}
with 
$a_k = 2 \ii (\cos\phi_k-h) -2 \gamma\sin\phi_k$.
Introducing 
$\aaa_{\pm k} = \frac{1}{\sqrt{2}} %
(\pm \frac{a_k}{\vert a_k\vert} \uu_{2k-1} + \uu_{2k})$
we diagonalize (\ref{eq:Hh2}) and obtain a 
quasi-particle hamiltonian 
\begin{equation}
\Hh = \sum_{k=1}^{n} \epsilon_k (\aaa_k^\dagger \aaa_k^{} - %
                                 \aaa_{-k}^\dagger \aaa_{-k}^{})
\label{eq:Hh3}
\end{equation}
with eigenvalues $\epsilon_k = \vert a_k \vert$ given by
a dispersion relation
\begin{equation}
\epsilon_k \equiv \epsilon(\phi_k) = %
  2\sqrt{ (h-\cos\phi_k)^2 + \gamma^2 \sin^2\phi_k}.
\label{eq:dispersion}
\end{equation}
The OSEE is calculated by transforming $\uu_k$ back to $\ww_j$ which gives us 
the time evolution of the canonical maps $\ww_j(t)$ and therefore the evolution 
matrix $\mathbf{\Phi}$,
\begin{equation}
\begin{pmatrix}\ww_{2j-1}(t)\\ \ww_{2j}(t)\end{pmatrix} = \sum_{l=1}^{n} 
\begin{pmatrix} f_{l-j}(t) & g_{l-j}(t) \\ %
              -g_{j-l}(t) &  f_{l-j}(t)  \end{pmatrix}^T
\begin{pmatrix} \ww_{2l-1} \\ \ww_{2l} \end{pmatrix}
\label{eq:wjtfourier}
\end{equation}
where $f_j, g_j$ are {\em real} functions given as Fourier series
\begin{equation}
\begin{pmatrix} f_j(t) \\ g_j(t) \end{pmatrix} %
= \frac{1}{n} \sum_{k=1}^{n} \ee^{-\ii j \phi_k} 
\begin{pmatrix} %
\cos\epsilon_k t \\ %
-\ii \frac{a_k}{\epsilon_k} \sin\epsilon_k t %
\end{pmatrix}.
\end{equation}

\begin{figure}
\centering
\includegraphics[width=0.9\columnwidth]{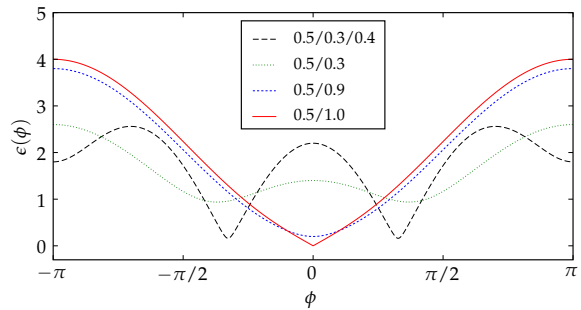}
\caption{Dispersion relation $\epsilon(\phi)$ (eq.~\ref{eq:dispersion}) for XY model with $\gamma=0.5$ and various 
$h$ (indicated in the figure) and for the modified model (\ref{eq:additional}) (long-dashed curve).}
\label{fig:dispersion}
\end{figure}
Comparing the dispersion relation (Fig.\ref{fig:dispersion}) 
for $(\gamma,h)$ with the OSEE (Fig.~\ref{fig:fOSEE})
we make the following
\\ 
{\bf Conjecture:}
\emph{
The prefactor $c$ in the logarithmic growth (\ref{eq:logOSEE}) 
of OSEE for infinite-index operators and open boundaries 
is given by the number $m$ of stationary points of the 
quasiparticle dispersion relation $\epsilon(\phi)$, as $c = m/6$,
if all stationary points are non-degenerate\footnote{%
A triple degenerate stationary point $\phi=0$ at $h=h_c$, $\gamma\neq 1$ 
would contribute  $1/12$ to the coefficient $c$.}.
}
\\
We observe that the nontrivial stationary points where
$\epsilon'(\phi)=0$ only exist when $\vert h\vert \leq \vert 1-\gamma^2\vert$ 
which determines the magnitude of the critical field separating the regions of $m=2$ and $m=4$,
\begin{equation}
h_{\rm c} = \vert 1-\gamma^2 \vert.
\label{eq:critical}
\end{equation}
We verify the conjecture by considering more general Wigner-Jordan solvable 
spin chains, say by adding a term 
$\sigma_j^{\rm y}\sigma_{j+1}^{\rm z}\sigma_{j+2}^{\rm y}$
-- or $\ii w_{2j-1}w_{2j+4}$ expressed with Majorana operators --
to the XY-hamiltonian $H$ in (\ref{eq:HXY}),
\begin{equation}
H' = H + \mu \sum_j \sigma_j^{\rm y}\sigma_{j+1}^{\rm z}\sigma_{j+2}^{\rm y}.
\label{eq:additional}
\end{equation}
Choosing  $\mu=0.4$ we obtain a dispersion relation 
with $6$ stationary points (Fig.\ref{fig:dispersion}) 
which agrees with the result for the OSEE growth with 
the prefactor $6/6$ (Fig.\ref{fig:fOSEE}). 

%
%



The idea of stationary points is supported by studying disordered chains 
where the Fourier transformation -- which would result in the dispersion relation 
-- is not applicable. We introduce disorder of strength $\varepsilon$
to either $\gamma$ or $h$, as $\gamma_j = \gamma + \varepsilon_j$ or $h_j = h + \varepsilon_j$, where 
$\varepsilon_j\in[-\varepsilon,\varepsilon]$ are uniformly distributed random numbers.
Although we shall only present data for $h$-disorder, similar behavior has been observed also in the
other case.

\begin{figure}[!h]
\centering
\includegraphics[width=0.9\columnwidth]{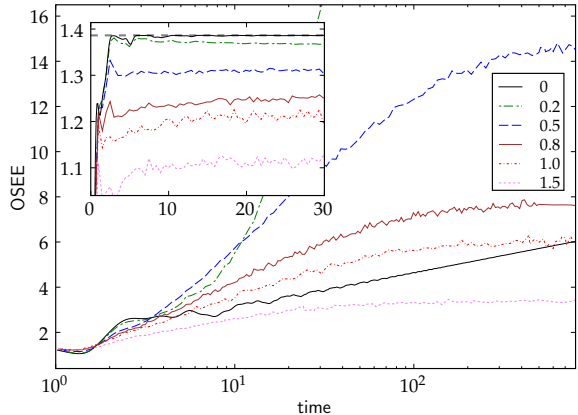}
\caption{OSEE for the XY model with $\gamma=0.2$ in a randomly 
disordered magnetic field $h_j \in [-\epsilon,\epsilon)$ 
averaged  over $1000$ realizations for disorder strengths
$\varepsilon \in \{0.05,0.2,0.5,0.8,1.0,1.5\}$.
While main figure corresponds to the infinite-index operator $F(t)$, 
the inset shows data for finite index operator $\sigma^z_{n/2}(t)$.
}
\label{fig:rndlarge}
\end{figure}

We first check for a finite-index initial operator $\sigma_{n/2}^{\rm z}$
for which we find that the OSEE saturates in time, both in homogeneous and disordered cases
(inset of Fig.~\ref{fig:rndlarge}). 
This result is expected since for any finite-index 
initial product operator $A$ (product of $w_j$'s) OSEE is upper-bounded by $S(t) \leq K \log 2$ 
where $K=\sum_{k=1}^{2n}\xbra{A}\ww_k^\dagger\ww_k^{}\xket{A}$ is the Majorana
index. The upper bound applies to \emph{any} Hamiltonian quadratic in 
Majorana operators and can be derived by a straight-forward generalization of
the proof for the critical quantum transverse Ising chain 
in \cite{prosenpra76}. However, no statement can be made on whether the OSEE 
for finite-index operators in the disordered model is higher or lower than
in the corresponding non-disordered case.

The OSEE of infinite-index initial operators is not bounded in general and the 
disorder significantly affects the dynamics of entanglement.
Again, we will restrict our interest to the
simplest infinite-index operator $F=w_1 w_2\cdots w_n$ 
as the results are qualitatively similar also for e.g. $\sigma_{n/2}^{\rm x}$
or $\sigma_{n/2}^{\rm y}$.
For a weak disorder, three stages can be identified  
in the evolution of infinite-index operators
(Fig.~\ref{fig:rndlarge}). 
Up to the time proportional to $1/\varepsilon$,
the OSEE roughly follows the non-disordered case, after which it 
grows with a rate proportional to $\varepsilon$ until it finally 
saturates to a plateau.
%
The saturation phenomenon is a consequence of 
Anderson localization of eigenvectors of $\Hh$ in the Majorana space
and must be contrasted with the logarithmic growth of the OSEE in time
for the corresponding non-disordered model.
The plateau value of OSEE  decreases with 
$\varepsilon$ although the quantitative relation cannot be established at present.

We have made an additional test where the external field 
is not randomly disordered but periodic such as 
$h_j = h + (-1)^j h'$.
Due to periodicity of $h_j$ we can again establish a dispersion relation which 
agrees with the logarithmic growth of the OSEE with the prefactor 
determined from the number of stationary points.


In conclusion, we have demonstrated a discontinuous transition in the quantum XY-model at the critical field strength
$h_{\rm c} = \vert 1-\gamma^2\vert$ which separates two phases of different entanglement production rate in the 
temporal dynamics of operators. The transition exactly corresponds to a far-from-equilibrium phase transition
in the steady state of the open XY chain \cite{prosenprl101}. Beyond the XY model, a general relation has been conjectured
which connects the prefactor of the {\em logarithmic} entanglement growth to the number of stationary points of
the quasiparticle dispersion relation. Interestingly, this {\em temporal scaling} of the operator space entanglement entropies is strongly reminiscent of 
the {\em size scaling} of the ground state entanglement entropy found for the critical (gapless) XY models \cite{its,keating}, where the
prefactor $c=m/6$ is related to the number of sign changes $m$ of the so-called symbol of an appropriate Toeplitz determinant.
However, in our context the logarithmic growth (versus the saturation) of the entanglement entropy is not determined by
gapless (gapped) nature of the Hamiltonian, but rather by the infinite (finite) index of the initial operators.

We acknowledge support by the grant P1-0044 of the Slovenian Research Agency.


\begin{thebibliography}{10}

\bibitem{cluster} R.~Raussendorf and H.-J. Briegel, Phys. Rev. Lett. {\bf 86},
5188 (2001).

\bibitem{vidal}
G.~Vidal, Phys. Rev. Lett. \textbf{91}, 147902 (2003); 
G.~Vidal, Phys. Rev. Lett. \textbf{93}, 040502 (2004).

\bibitem{white}
S. R. White,  Phys. Rev. Lett. \textbf{69}, 2863 (1992).

\bibitem{latorre}
G.~Vidal, J.~I.~Latorre, E.~Rico, and A.~Kitaev, 
   Phys. Rev. Lett. {\bf 90}, 227902 (2003);
J.~I.~Latorre, E.~Rico, and G.~Vidal, Quant.~Inf.~Comp.{\bf 4}, 48 (2004).

\bibitem{hastings}
M.~B.~Hastings, J.~Stat. Mech. P08024 (2007);
M.~B.~Hastings, Phys. Rev. B \textbf{76}, 035114 (2007).

\bibitem{verstraeteprb73}
F.~Verstraete and J.~I.~Cirac, Phys. Rev. B \textbf{73}, 094423 (2006).

\bibitem{prosenpre75}
T. Prosen and M. {\v Z}nidari{\v c}, Phys. Rev. E \textbf{75}, 015202(R) (2007).

\bibitem{verstraeteprl93}
F.~Verstraete, J.~J.~Garcia-Ripoll, and J.~I.~Cirac,
Phys. Rev. Lett. \textbf{93}, 207204 (2004).

\bibitem{calabreseJSM4}
P. Calabrese and J. Cardy, J. Stat. Mech. Theor. Exp. P04010 (2005).

\bibitem{znidaricpra}
M. \v{Z}nidari\v{c}, T. Prosen, and I. Pi\v{z}orn, 
Phys. Rev. A \textbf{78}, 022103 (2008).

\bibitem{pleniohartmann}
M.~J.~Hartmann, J.~Prior, S.~R.~Clark, and M.~B.~Plenio, 
Phys. Rev. Lett. \textbf{102}, 057202 (2009).


\bibitem{prosenpra76}
T.~Prosen and I.~Pi\v{z}orn, Phys. Rev. A~\textbf{76}, 032316 (2007).

\bibitem{zanardi}
P.~Zanardi, Phys. Rev. A \textbf{63}, 040304(R) (2001);
X.~Wang and P.~Zanardi, Phys. Rev. A~\textbf{66}, 044303 (2002);
X.~Wang, B.~C.~Sanders, and D.~W.~Berry,
Phys. Rev. A \textbf{67}, 042323 (2003);
M.~A.~Nielsen {\it et al}, Phys. Rev. A \textbf{67}, 052301 (2003);
K.~{\.Z}yczkowski and I.~Bengtsson, Open Syst. Inf. Dyn. \textbf{11}, 3-42 (2004).

\bibitem{prosen3Q}
T.~Prosen, New~J. Phys.~\textbf{10}, 043026 (2008).

\bibitem{prosenprl101}
T.~Prosen and I.~Pi\v{z}orn, Phys. Rev. Lett.~\textbf{101}, 105701 (2008).

\bibitem{its}
A.~R.~Its, B.-Q.~Jin and V.~E.~Korepin, 
J. Phys. A: Math. Gen. \textbf{38}, 2975 (2005); 
arXiv:quant-ph/0606178.

\bibitem{keating}
J.~P.~Keating and F.~Mezzadri, Phys. Rev. Lett. {\bf 94}, 050501 (2005):

\end{thebibliography}
\end{document}